\documentclass[reprint,aps, superscriptaddress,nofootinbib]{revtex4-1}

\usepackage{graphicx}
\usepackage{dcolumn}
\usepackage{bm}

\usepackage{amsmath}
\usepackage{amssymb}
\usepackage{amsfonts}
\usepackage{amsthm}
\usepackage{color}
\usepackage{subfig}
\usepackage{MnSymbol}
\begin{document}

\title{Deconstructing symmetry breaking dynamics}

\date{\today}

\author{Fumika Suzuki}
\email{fsuzuki@lanl.gov}
 \affiliation{%
Theoretical Division, Los Alamos National Laboratory, Los Alamos, New Mexico 87545, USA
}
 \affiliation{%
Center for Nonlinear Studies, Los Alamos National Laboratory, Los Alamos, New Mexico 87545, USA
}

\author{Wojciech H. Zurek}

 \affiliation{%
Theoretical Division, Los Alamos National Laboratory, Los Alamos, New Mexico 87545, USA
}

  \begin{abstract}
The Kibble-Zurek mechanism (KZM) successfully predicts the density of topological defects deposited by the phase transitions, but it is not clear why. Its key conjecture is that, near the critical point of the second-order phase transition, critical slowing down will result in a period when the system is too sluggish to follow the potential that is changing faster than its reaction time. The correlation length at the freeze-out instant $\hat t$ when the order parameter catches up with the post-transition broken symmetry configuration is then decisive, determining when the mosaic of broken symmetry domains locks in topological defects. To understand why the KZM works so well we analyze Landau-Ginzburg model and show why temporal evolution of the order parameter plays such a key role. The analytical solutions we obtain suggest novel, hitherto unexplored, experimentally accessible observables that can shed light on symmetry breaking dynamics while testing the conjecture on which the KZM is based.
 \end{abstract}

\maketitle

The Kibble–Zurek mechanism (KZM) combines Kibble’s insight \cite{kibble,kibble2} that, during cosmological phase transitions, the causal independence of distinct Hubble volumes leads to independent choices of broken symmetry and thus to the formation of topological defects, with the realization \cite{whz,whz2,whz3} that the universality class and associated critical exponents determine how the defect density scales with the quench rate in a second-order phase transition. This scaling is based on a conjecture about the dynamics of the order parameter near the critical point: It uses equilibrium properties of the transition, such as the critical scaling of the relaxation time and the healing length, to predict a nonequilibrium outcome --- the relic density of topological defects formed after the transition. The ``freeze-out" (or ``adiabatic-impulse-adiabatic" approximation) and the ``sonic horizon" paradigms explain the defect density by focusing on different aspects of the order parameter dynamics in the near-critical region. Despite their differences, both approaches lead to the same scaling law. We deconstruct the order parameter dynamics in the Landau-Ginzburg model of the phase transition to identify the origin of the KZM scaling.

\begin{figure}[t]
\centering
\includegraphics[width=0.9\linewidth]{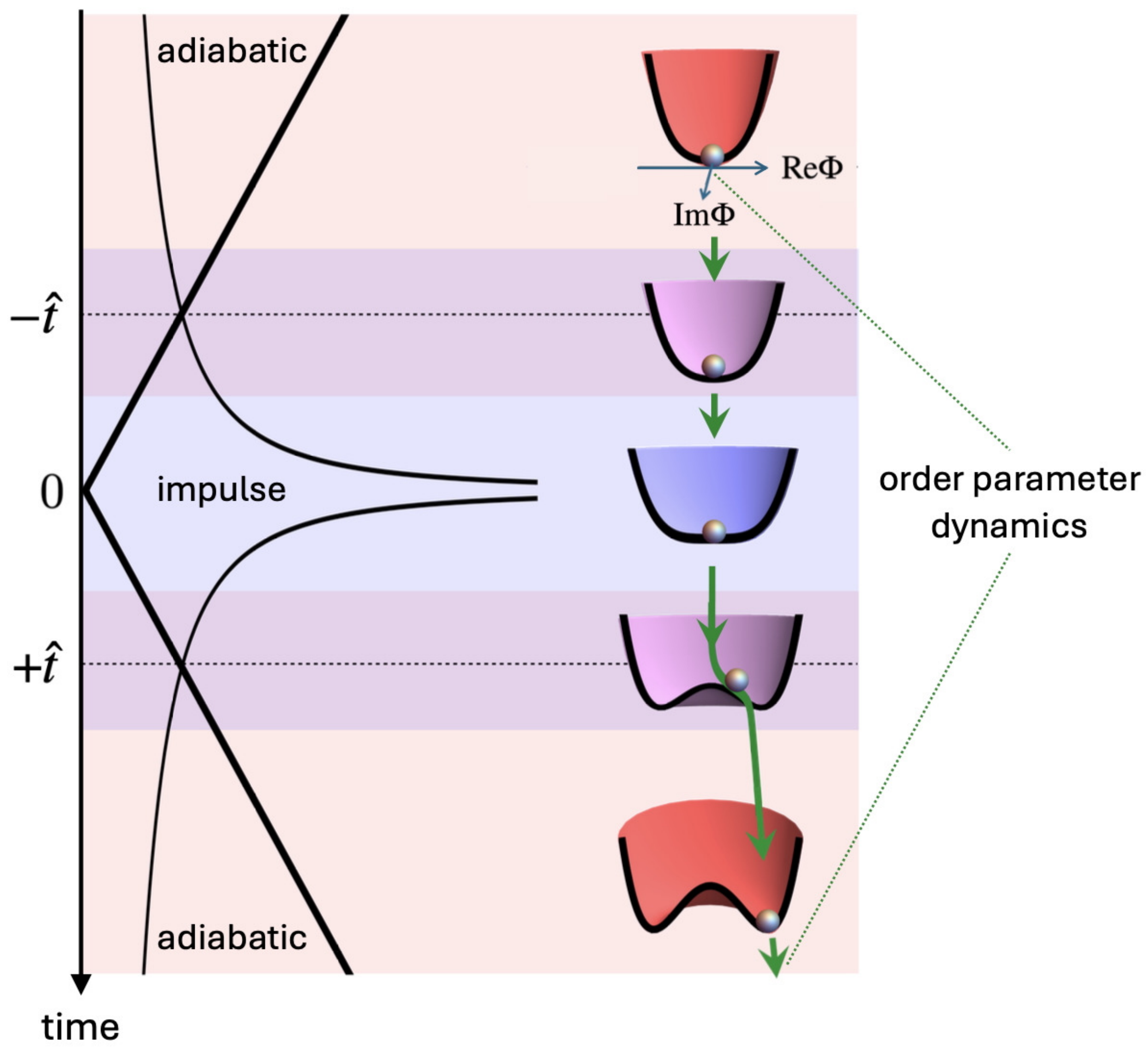}
\caption{Symmetry breaking dynamics  in a second order phase transition. Landau-Ginzburg model attributes symmetry breaking to the change of the thermodynamic potential from the single-minimum parabola (top) to the degenerate minima (bottom). When the order parameter is a real field (case represented by the dark line above) there are just two broken symmetry minima. When the field is complex (i.e., order parameter in a superfluid), the potential assumes a ``sombrero’’ shape, with continuum of the broken symmetry states.  We investigate how the temporal evolution of the order parameter (which starts for $t<0$ in the single symmetric minimum, but must choose one of the broken symmetry possibilities after the transition, see Fig. \ref{fig3}) leads to the mosaic of domains that lock in topological defects. The Kibble-Zurek mechanism posits that the size of these domains is set by the correlation length at the instant $+\hat t$ when the order parameter catches up with the new broken symmetry minimum. We show that this temporal evolution can be analyzed using ordinary differential equations with solutions that mark the freeze-out instant $+\hat t$.}
\label{fig1}
\end{figure}

Critical slowing down plays the key role. The relaxation time diverges in the vicinity of the critical point with vanishing $\epsilon$, the dimensionless distance from the critical point: 
$$ \tau(\epsilon(t)) \cong \tau_o / |\epsilon(t)|^{\nu z} . $$
Above, $\nu$ and $z$ are critical exponents. The freeze-out paradigm compares it with the timescale $\epsilon / \dot \epsilon$ on which the distance from the critical point varies during the quench that can be induced, e.g., by the change of pressure or temperature. In the narrow near-critical range linear dependence on time: 
$$ \epsilon(t) = t / \tau_Q $$
with $\tau_Q$ the quench timescale can be usually assumed. Hence;
$$\epsilon / \dot \epsilon = t .$$
Consequently, relaxation time $\tau(\epsilon)$ and $\epsilon / \dot \epsilon $ (i.e., time from the critical point)  equal at the instant $t=\pm \hat t$ when $\tau_o / |\hat t/\tau_Q|^{\nu z}= \hat t $. Thus:
$$\hat t = (\tau_o \tau_Q^{\nu z})^{\frac 1 {1+\nu z}} ,$$
and;
$$ \hat \epsilon = \hat t / \tau_Q = (\frac {\tau_o} {\tau_Q})^{\frac 1 {1+\nu z}} .$$
The KZM asserts that the evolution of the order parameter will cease to be adiabatic at $t\approx-\hat t$ before the critical point, enter an impulse regime where the changes imposed by the quench outpace relaxation, 
and will return to adiabatic after $t=+\hat t$, but with the initial conditions set by the fluctuations introduced before $+\hat t$. 

To estimate defect density one recognizes \cite{whz,whz2,whz3} that at the freezeout time $\hat t$ the correlation length is:
$$ \hat \xi = \xi_o/|\hat \epsilon|^{\nu}= \xi_o \left(\frac {\tau_Q} {\tau_o}\right)^{\frac \nu {1+\nu z}} .$$
A single ``unit’’ of topological defect (e.g., a single monopole, a section of the line defect, or a ``tile'' of a domain wall) is then expected to appear in a domain of size $\hat \xi$. This leads to the estimated dependence of the defect density on the quench time $\tau_Q$:
$$ {\cal N} \simeq {\hat \xi}^d / {\hat \xi}^D = \xi_o^{(d-D)} \left(\frac {\tau_Q} {\tau_o}\right)^{(d-D) {\frac {\nu } {1 +\nu z}}}.$$
Above, $d$ is the dimensionality of defects ($d=0$ for monopoles, $d=1$ for string-like flux lines, and $d=2$ for domain walls), and $D > d$ is the dimensionality of the space, while $\xi_o$ and $\tau_o$ are determined by microphysics.

The sonic horizon paradigm focuses instead on the growth of broken symmetry domains from the preexisting fluctuations that act as seeds. The relevant sound velocity with which the newly selected broken symmetry vacuum spreads is expected to be:
$$ u =  \xi (\epsilon) / \tau (\epsilon) = (\xi_o/\tau_o) | \epsilon|^{\nu(z-1)}. $$
When the front of the new broken symmetry phase propagates with this velocity over the time interval $\sim \hat t$, the size of the domains that choose the same broken symmetry scales as $\hat \xi = \xi_o (\frac {\tau_Q} {\tau_o})^{\frac \nu {1+\nu z}}$, as in the freeze-out paradigm. The defect density ${\cal N}$ derived above also follows.

Both paradigms are discussed in \cite{whz,whz2,whz3,kzm}, where it is emphasized that the crucial prediction of the KZM is the scaling of the post-transition defect density with the quench time. Although the physical pictures invoked by the two paradigms differ, they can nevertheless peacefully coexist in both quantum and classical settings \cite{sonic}. This coexistence, and especially the fact that they yield essentially the same scaling, is surprising, as there are transitions that rely on different, non-sonic modes of propagation of the broken symmetry phase front. For instance, transitions involving conserved order parameter \cite{damski2007dynamics,lamacraft2007quantum} or diffusive spreading of the new phase \cite{mono, lin} do not fit the sonic horizon scenario, and neither does structure formation in steady state (rather than equilibrium) settings \cite{ducci1999order, casado2006testing} such as Rayleigh-Benard instability. There even exists an example \cite{damski} where spatial degrees of freedom are absent.

We conclude that the temporal evolution encapsulated in the freeze-out paradigm captures the essential dynamics of phase transitions relevant for defect formation. Our aim is to understand why this is the case. We will show that the temporal evolution of the order parameter plays a central role and captures the essence of the KZM. Moreover, our results highlight a new observable that has so far remained largely unexplored but should be accessible experimentally: the quench-induced temporal evolution of the order parameter in the near-critical region (Fig. \ref{fig1}).

\section*{Deconstructing the Kibble-Zurek mechanism: the temporal evolution}

\begin{figure}[t]
\centering
{%
\includegraphics[clip,width=0.8\columnwidth]{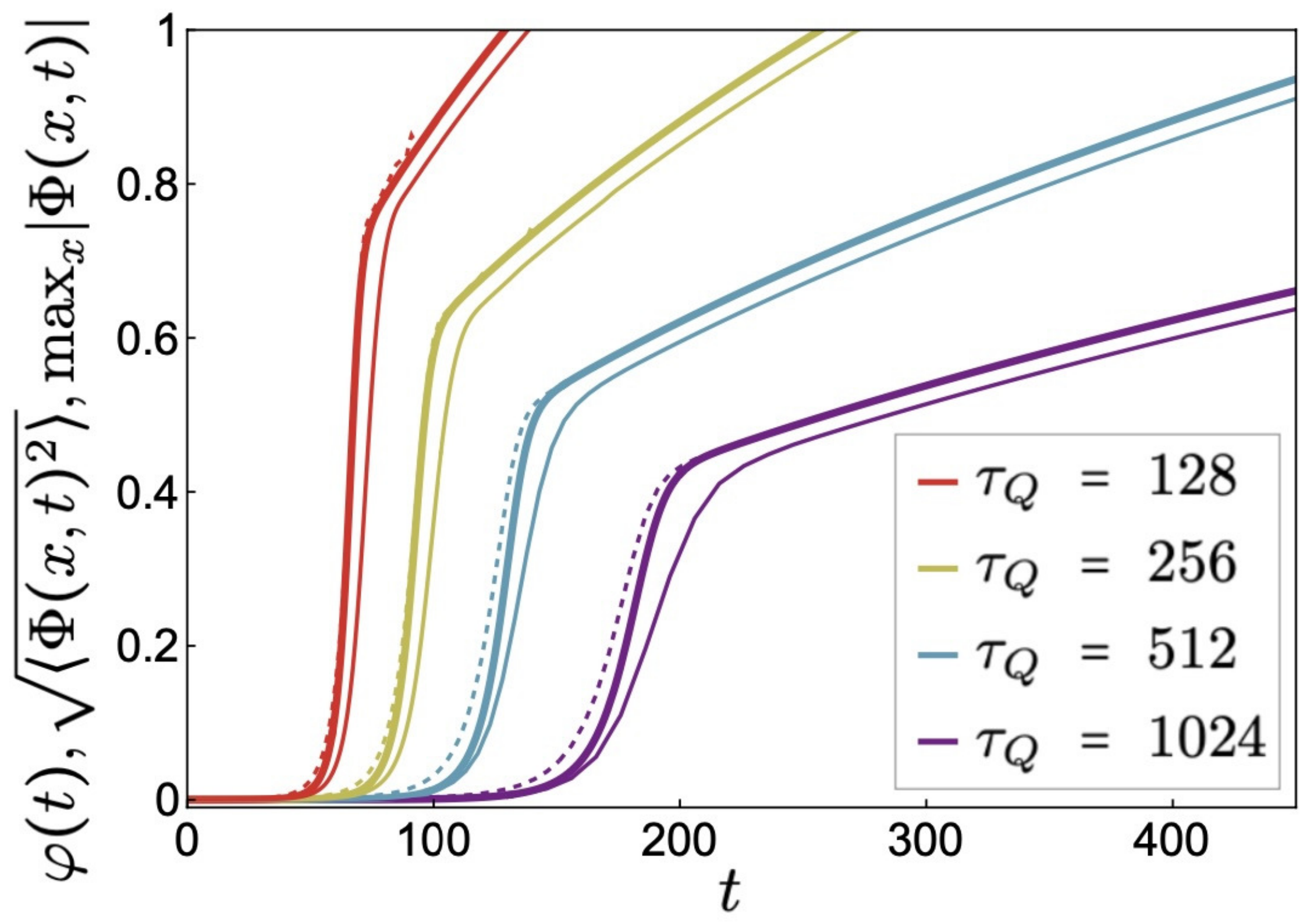} 
}
\caption{Time evolution of the order parameter: comparison between the analytical solution of the ordinary differential equation (\ref{ode1}) and the numerical solution of the Langevin equation for a real scalar field in (1+1) dimensions [\ref{langevin}]. The thick lines represent the analytical solutions $\varphi (t)$ (Eq. (\ref{solu})) with $\eta=1$ and $\varphi (0)=10^{-4}$  for various quench timescales $\tau_Q$, while the numerical results $\sqrt{\langle \Phi (x,t)^2\rangle}$ (thin lines), $\displaystyle\mbox{max}_x |\Phi (x,t)|$ (dashed lines) are obtained by solving  Eq. (\ref{langevin})  with $\eta=1$ and $\theta=10^{-8}$. From left to right, $\tau_Q=128,256,512,1024$ respectively.} 
\label{fig2}
\end{figure}

The usual ``workhorse'' in the study of the nonequilibrium dynamics during symmetry-breaking phase transitions is a Langevin partial differential equation (PDE) with a time-dependent Landau-Ginzburg potential,  $V(\Phi)=(\Phi^{4}-2\epsilon (t) \Phi^{2})/8$,  for the real scalar field $\Phi$ representing the order parameter \cite{lg,lg2,vortex8,vortex9, jacek2,fumika}:
\begin{eqnarray}\label{langevin}
\ddot{\Phi} (\mathbf{x},t)+\eta \dot{\Phi}(\mathbf{x},t) -\nabla^2 \Phi (\mathbf{x},t) +\partial_{\Phi} V(\Phi) =\vartheta (\mathbf{x},t)
\end{eqnarray}
where  the noise term $\vartheta$ has correlations given by;
$$\langle \vartheta (\mathbf{x},t), \vartheta  (\mathbf{x}',t')\rangle = 2\eta \theta \delta (\mathbf{x}'-\mathbf{x})\delta (t'-t).$$ 
Here, $\eta$ represents the damping constant, and $\theta$ is the temperature of the reservoir.

We now ``deconstruct'' the Langevin PDE used to study defect formation, and consider temporal evolution separately from its spatial part and the noise. As we shall see, the resulting dynamics of the order parameter suggests the succession of the adiabatic-impulse-adiabatic stages. Distinguishing between these stages helps explain why the instants $- \hat t$ and $+ \hat t$, which mark the transitions between them, are crucial for establishing the KZM scaling. 

\begin{figure}[t]
\centering
{%
\includegraphics[clip,width=0.98\columnwidth]{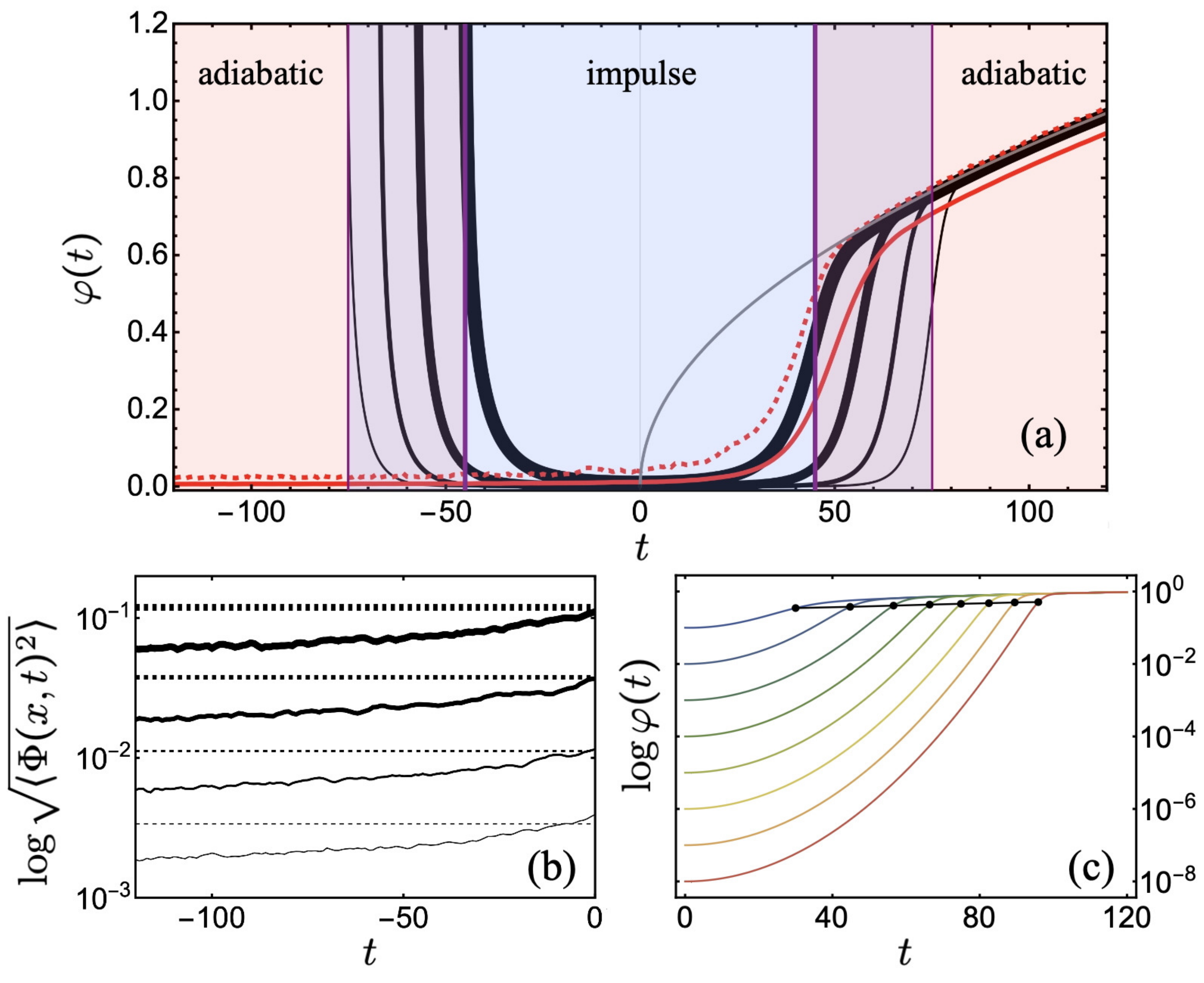} 
}
\caption{Order parameter dynamics for varying noise temperature $\theta$ and initial conditions $\varphi (0)$. (a) The order parameter $\varphi(t)$ from Eq. (\ref{solu}) for $\eta=1$, $\tau_Q=128$, and $\varphi(0) = 10^{-2}$, $10^{-3}$, $10^{-4}$, $10^{-5}$ (thick to thin lines), compared with numerical results: $\sqrt{\langle \Phi^2 \rangle}$ (solid red) and $\max_x |\Phi|$ (dashed red) from Eq. (\ref{langevin}) for a real scalar field in (1+1) dimensions with $\eta=1$, $\theta=10^{-4}$ (see Fig. \ref{fig2}). The gray line shows equilibrium $|\varphi_{\rm min}| = \sqrt{\epsilon}$. Vertical purple lines mark $\pm \hat{t}$ for $\varphi(0)=10^{-2}$ (thick) and $10^{-5}$ (thin).
(b) Noise-induced $\sqrt{\langle \Phi^2 \rangle}$ with $\theta= 10^{-2}$ to $10^{-5}$ (thick to thin) slowly accumulates before $t = 0$, setting $\varphi(0)\approx \sqrt{\langle \Phi (x,t=0)^2\rangle} \approx\frac12 \sqrt{2\theta /\hat{\epsilon}}$ (dashed lines) for rapid post-transition growth.
(c) Plots of $\log \varphi(t)$ for $\varphi (0)=10^{-1} ...10^{-8}$.  As $\varphi (t) \approx {\varphi (0)} e^{t^2/4\eta\tau_Q}$ when $t \in [-\hat{t},+\hat{t} ]$, order parameter buffeted by noise is amplified, but the freeze-out time $+\hat{t}$ depends only logarithmically on $\varphi(0)$, making it  insensitive to its precise value, and, hence, insensitive to $\theta$, the noise temperature.}
\label{fig3}
\end{figure}

We focus on the ordinary differential equation (ODE) in which $\varphi (t)$ depends solely on time:
\begin{eqnarray}\label{ode1}
\ddot{\varphi}(t)+\eta \dot{\varphi}(t)=-\partial_{\varphi}V(\varphi(t))
\end{eqnarray}
obtained from Eq. (\ref{langevin}) by omitting the spatial degrees of freedom and the noise term.
It captures the essence of the temporal behavior of the order parameter $\varphi (t)$ without the need to rely on the full PDE (Eq. (\ref{langevin})). Indeed, in the overdamped case where $\ddot{\varphi}(t) \ll \eta \dot{\varphi}(t)$, Eq. (\ref{ode1}) is a Bernoulli ODE, 
\begin{eqnarray}
\eta \dot{\varphi}(t)+\frac12 (\varphi (t)^3-\epsilon (t) \varphi (t))=0
\end{eqnarray}
which can be solved analytically:
\begin{eqnarray}\label{solu}
\frac{\varphi (t)}{\varphi (0) }=\frac{ e^{t^2/4\eta\tau_Q}}{\sqrt{1+\varphi (0)^2\sqrt{\frac{\pi \tau_Q}{2\eta}}\mathrm{erfi}(\frac{t}{\sqrt{2\eta\tau_Q}})}}
\end{eqnarray}
where $\varphi (0) =\varphi (t=0)$ and $\mathrm{erfi}(y)=\frac{2}{\sqrt{\pi}}\int_0^y e^{y'^2}dy'$. This function predicts the evolution of the order parameter in symmetry-breaking phase transitions that can be modeled by Landau-Ginzburg theory. Indeed, the time evolution of the order parameter given by numerical simulation of the Langevin PDE, Eq. (\ref{langevin}) in (1+1) dimensions, is in close agreement with the analytical solution of $\varphi (t)$ as shown in Fig. \ref{fig2}. Thus, $\sqrt{\langle \Phi (x,t)^2\rangle}$ (thin lines) obtained by averaging over the spatial dependence of $\Phi (x,t)^2$ is somewhat smaller than the solution of Bernoulli ODE given by Eq. (\ref{solu}) (thick lines), as its value is suppressed by the presence of topological defects. Maximum values of $|\Phi (x,t)|$, $\displaystyle\max_x |\Phi(x,t)|$ (dashed lines), by contrast, are slightly larger than $\varphi (t)$, enhanced by the random walk due to noise. Therefore,
the evolution of the order parameter $\varphi$ is largely captured by solving the ODE  Eq. (\ref{ode1}). As that temporal equation is the same in multi-dimensional spaces, we expect that our conclusions also follow when there is more than one spatial dimension.

In the pre-transition symmetric phase, noise leads to fluctuations around $\varphi = 0$, as shown in Fig. \ref{fig3}. This behavior is, of course, not observed when solutions of the noise-free Eq. (\ref{ode1}) are traced back to the pre-freezeout epoch, $t \le -\hat t$, since—consistent with the Bernoulli equation—the order parameter can attain a finite and modest value at $t = 0$ only if it starts with a suitably large $|\varphi(t)|$ in the distant pre-transition past (Fig \ref{fig3} (a)). By contrast, in numerical solutions of the Langevin PDE, the accumulation of noise results in a modestly fluctuating $\varphi(t)$ in the near-critical regime. The influence of the noise is thus effectively encoded in the ``initial condition" $\varphi (0)$, and the resulting
dependence of $\varphi (0)$ on $\theta $ can be estimated: Before $t=0$, the potential is a harmonic $V_{\rm har} (\Phi)=-\frac{1}{4}\epsilon (t) \Phi^2$ near $\Phi = 0$.  Since the temperature $\theta$ corresponds to the kinetic energy of $\Phi$, we can estimate $\sqrt{\langle \Phi^2\rangle}\approx \frac12 \sqrt{2\theta/(-\epsilon (t))}$ where the factor 1/2 results from averaging over $x$ (Fig. \ref{fig3} (b)). More precise estimate is not really needed for our purpose, as the all-important $\hat t$ depends only logarithmically on $\varphi (0)$ (and, hence, on $\sqrt{\langle \Phi (x,t=0)^2\rangle}$) (Fig. \ref{fig3} (c)).

\section*{Evolution of the Gross-Pitaevskii wave function}

\begin{figure}[t]
\centering
{%
\includegraphics[clip,width=0.8\columnwidth]{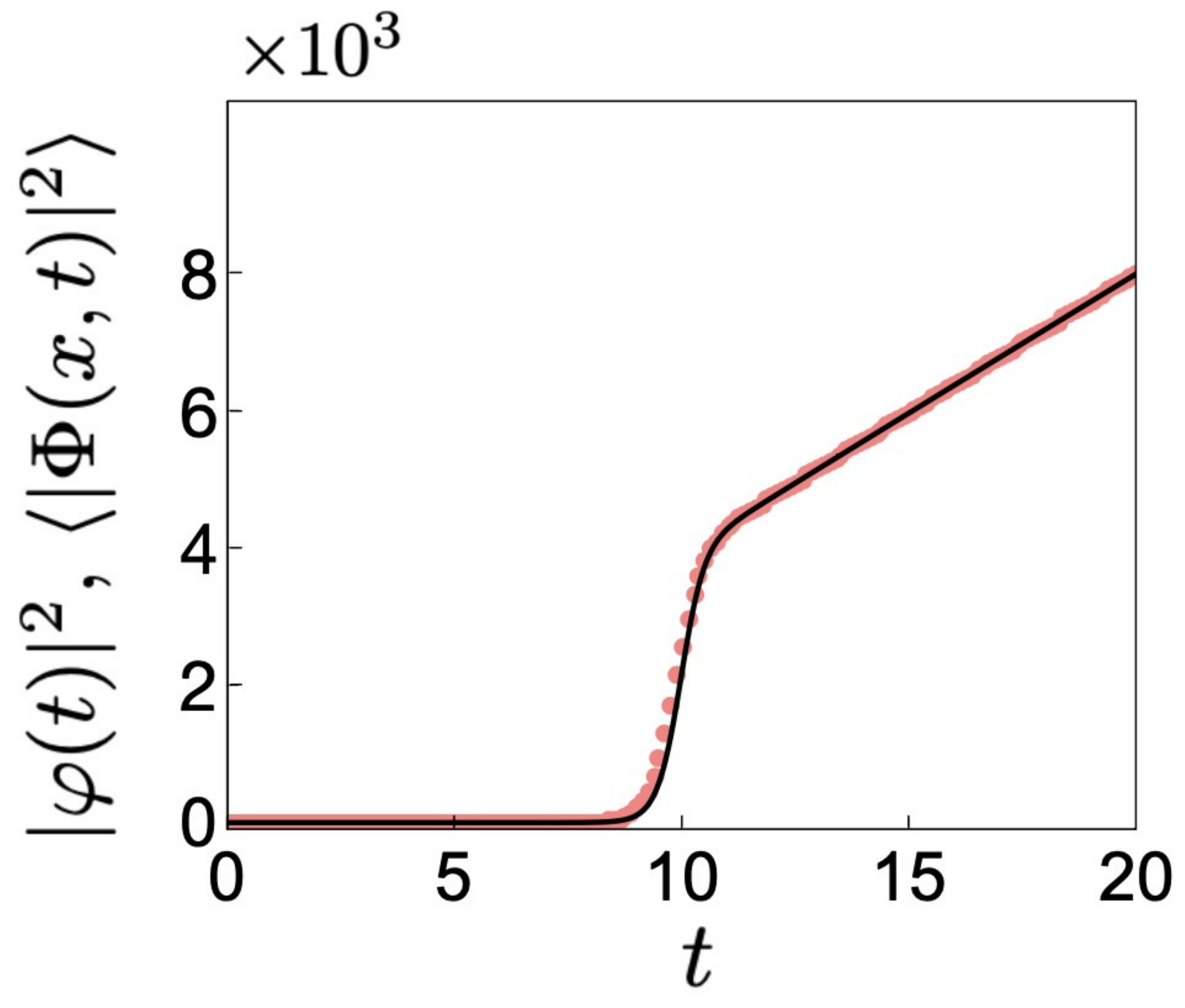} 
}
\caption{Order parameter evolution for the Gross–Pitaevskii equation. The black line represents the analytical solution for the condensate number density $|\varphi (t)|^2$ (Eq. (\ref{solu2})) with  $\varphi (0)=3\times 10^{-3}$, while the corresponding numerical result $\langle |\Phi (x,t)|^2\rangle$ (light red dots) is obtained by solving the stochastic Gross–Pitaevskii equation Eq. (\ref{gpe}) in (1+1) dimensions for a complex wave function-like order parameter with $\theta=10^{-3}$. For both plots, $\gamma=10^{-2}$, $g=0.05$, and quench timescale $\tau_Q=0.05$.} 
\label{fig4}
\end{figure}

This approach also applies to a complex wave function-like  order parameter $\Phi(\mathbf{x},t)$, such as in the stochastic Gross–Pitaevskii equation, which is commonly used to describe Bose–Einstein condensates \cite{bec,bec2,bec3,stgpe,stgpe2,das} and can be written as
\begin{eqnarray}\label{gpe}
(i-\gamma) \dot{\Phi}(\mathbf{x},t)&=& -\frac12 \nabla^2 \Phi (\mathbf{x},t) -\epsilon (t)\Phi (\mathbf{x},t) \nonumber\\
&&+g |\Phi (\mathbf{x},t)|^2\Phi (\mathbf{x},t) +\vartheta (\mathbf{x},t)
\end{eqnarray}
where $\gamma$ represents the dissipation and $g$ is the non-linearity parameter. 

In this case as well, the equation can be deconstructed into a temporal ODE by omitting the spatial degrees of freedom and the noise term:
\begin{eqnarray}
(i-\gamma) \dot{\varphi}(t)=-\epsilon (t) \varphi (t)+g|\varphi (t)|^2\varphi (t).
\end{eqnarray}
Applying the ansatz $\varphi =|\varphi (t)| e^{i\Theta (t)}$ to the condensate wavefunction, this equation also yields an analytically solvable form of the Bernoulli ODE for $|\varphi (t)|$:
\begin{eqnarray}
(1+\gamma^2)\frac{d |\varphi (t)|}{dt}=\gamma \epsilon (t)|\varphi (t)|-\gamma g |\varphi (t)|^3 .
\end{eqnarray}
Its solution can be written as
\begin{eqnarray}\label{solu2}
\frac{|\varphi (t)|}{|\varphi (0)| }=\frac{ e^{\gamma t^2/2(1+\gamma^2)\tau_Q}}{\sqrt{1+g\varphi (0)^2\sqrt{\frac{\pi \gamma \tau_Q}{1+\gamma^2}}\mathrm{erfi}(\frac{\sqrt{\gamma}t}{\sqrt{(1+\gamma^2)\tau_Q}})}}
\end{eqnarray}
where $\varphi (0)$ is  determined by the size of the noise.

Fig. \ref{fig4} shows a comparison between the numerical solution  of the stochastic Gross–Pitaevskii equation (Eq. (\ref{gpe})), representing the time evolution of the condensate number density $\langle |\Phi (x,t)|^2\rangle$ (spatially averaged), and that given by the analytical solution (Eq. (\ref{solu2}))\footnote{We note that in the simulations of the Gross-Pitaevskii equation ``truncated Wigner method’’ (see, e.g., \cite{blakie}) is often employed, with the noise introduced to the initial $\Phi(x,t)$. Thus, unlike in the case of the Langevin Eq. (1) (where noise is injected while the system evolves), quantum fluctuations at or before at $-\hat t$ would evolve into the seeds of topological defects at $+ \hat t$.}.

This result demonstrates that analytical functions as simple as Eqs. (\ref{solu}, \ref{solu2}) can capture surprisingly well the time evolution of the order parameter in symmetry-breaking phase transitions for broad range of systems  that can be modeled within the Langevin /  Landau-Ginzburg or Gross-Pitaevskii framework. The experimentally accessible time evolution of the order parameter in transitions involving magnetization, polarization or condensate density could be modeled using Eqs. (\ref{solu}, \ref{solu2}).

The KZM does not directly address the time evolution of the order parameter. Nevertheless, the temporal behavior of the order parameter we discussed anticipates its essential features. In the next section, we show how these features emerge and relate them to the core concepts of the mechanism. 

\section*{Evolution of the order parameter and the Kibble-Zurek mechanism}

The key insight \cite{whz,whz2,whz3} that leads to the KZM scaling is the realization that, near the critical point of the second-order phase transition, critical slowing down will result in a time interval $[-\hat{t},+\hat{t}]$ where the order parameter is too sluggish to adjust to the potential which is changing faster than its reaction time. Thus, while outside this interval the order parameter can be in approximate equilibrium, within the interval $[-\hat{t},+\hat{t}]$ its evolution
``cannot keep up'' with time-dependent $V(\Phi)$. This is of little consequence when $t<0$, as prior to symmetry breaking the order parameter only fluctuates around the symmetric vacuum. Nevertheless, fluctuations imparted near and especially after $-\hat t$ are unconstrained by $V(\Phi)$. They seed topological defects that germinate as $\Phi(x,t)$ begins to catch up when $t>0$ with the local broken symmetry minima of the potential. As a consequence, near $+\hat t$ defects become frozen in the broken symmetry configuration in ways that depend on the nature of the system \cite{sonic}.

The KZM scaling appears to be insensitive to the details of that evolution. 

This broad applicability of the KZM suggests that, as proposed in \cite{whz,whz3,whz2}, the critical slowing down is key to its success. 
We now return to Fig. \ref{fig3} to discuss the connection between the KZM and the dynamics of the order parameter. Fig. \ref{fig3} (a) shows the plot of the solution (Eq. (\ref{solu})) with $\eta=1$, $\tau_Q=128$ for several values of $\varphi (0)$. After $t=0$ all exhibit similar behavior: The initial period (where $\varphi (t)$ slowly increases till $+\hat t$) is followed by a ``jump" where the solution catches up with the local broken symmetry equilibrium, and thereafter follows 
the broken symmetry minimum of the Landau-Ginzburg potential. 

This behavior is suggested by the adiabatic-impulse-adiabatic scenario \cite{whz,whz2,whz3,kzm}, and was seen in numerical simulations \cite{adolfo, jump}.  The start of the rapid rise can be identified with the freeze-out time $+\hat{t}$, when the order parameter evolution switches from the sluggish pre-transition pace to catch up with the post-transition broken symmetry equilibrium value $\sim \sqrt{\epsilon(t)}$. Note that before $t=0$ (when the critical point is transversed) solution $\varphi(t)$ of Eq. (\ref{ode1}) ``jumps down'' to relatively small values in the impulse regime, reaching them at the instant suggestive of $-\hat{t}$. When $t\in [-\hat{t},+\hat{t}]$ evolution is slow compared to these two ``jumps". As was already discussed, noise plays a key role there.

Fig. \ref{fig3} (c) shows the plots of $\log(\varphi)$ for various $\varphi(0)$. Black circles indicate the freeze-out time $+\hat{t}$ for each $\varphi(0)$, where $\ddot{\varphi} = 0$. This inflection point signifies a ``jump" which we identify with $+\hat t$. It demonstrates that the instants $\pm \hat{t}$ depend only logarithmically on $\varphi(0)$ at the critical point that incorporates the effect of the noise.

\begin{figure}[t]
\centering
{%
\includegraphics[clip,width=0.9\columnwidth]{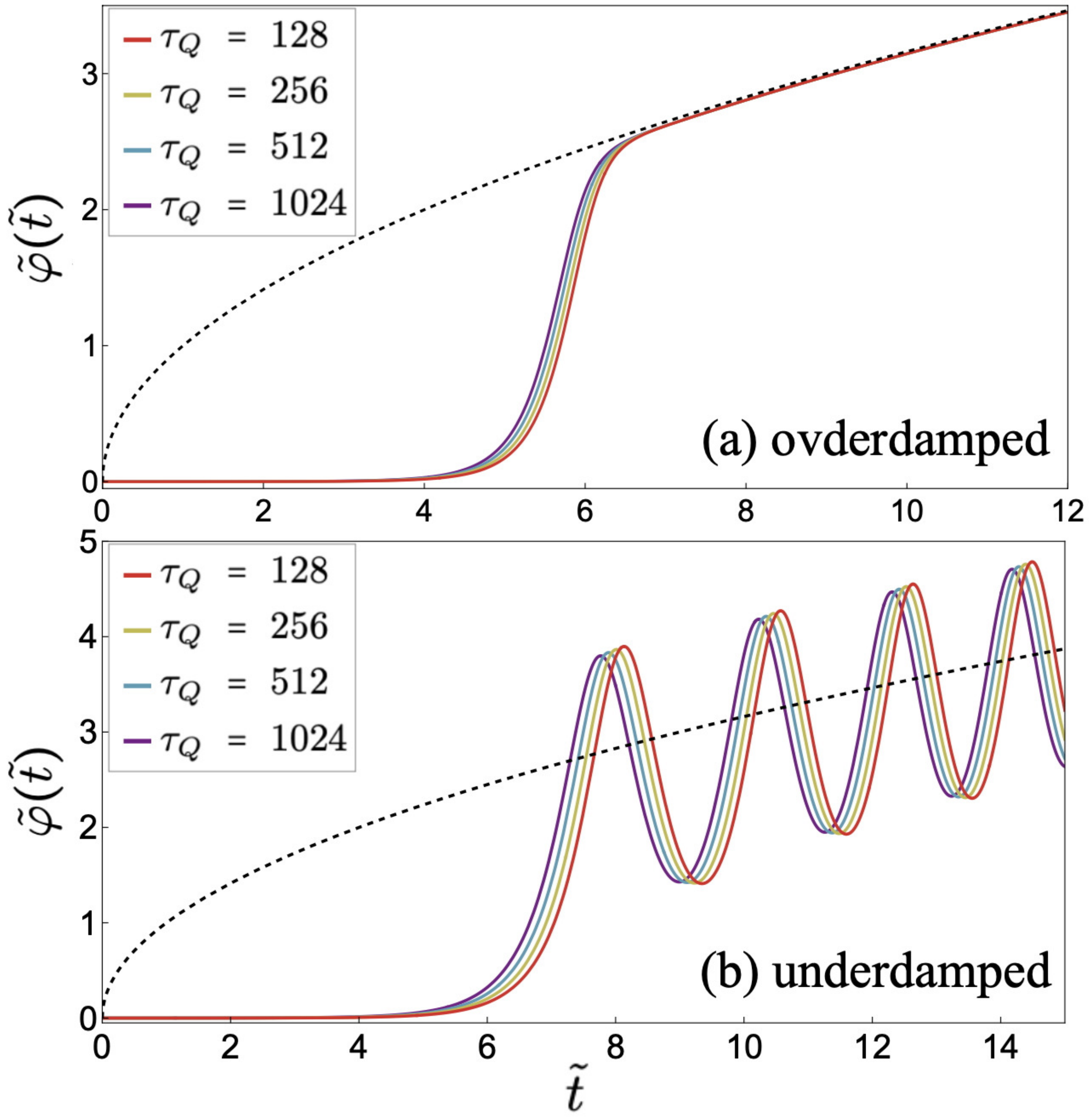} 
}
\caption{Order parameter dynamics given by Eq. (\ref{ode1}) in the overdamped and underdamped cases corresponding to a real scalar field described by the Langevin equation. The rescaled  solution $\tilde{\varphi} (\tilde{t})$ in the  overdamped case with $\eta=1$ where the first term $\ddot{\varphi}$ of Eq. (\ref{ode1})  is discarded (a) and in the  underdamped case where the second term $\eta \dot{\varphi}$ of Eq. (\ref{ode1}) is discarded (b). $\tilde{\varphi} = \sqrt[4]{\tau_Q/\eta}  \varphi$ and $\tilde{t} = t / \sqrt{\eta \tau_Q}$ in (a). $\tilde{\varphi}=\tau_Q^{1/3}\varphi$, $\tilde{t}= t/\tau_Q^{1/3}$ in (b).  $\varphi (0)=10^{-4}$ and various quench timescales $\tau_Q$. The dashed line represents $\tilde{\varphi}=\sqrt{\epsilon (\tilde{t})}$, the location of the broken symmetry minimum of the potential $V$.} 
\label{fig5}
\end{figure}

Plotting the rescaled solution $\tilde{\varphi}(\tilde{t})$, where $\tilde{\varphi} = \sqrt[4]{\tau_Q/\eta}  \varphi$ and $\tilde{t} = t / \sqrt{\eta \tau_Q}$, reveals a clear similarity among the curves for different quench timescales $\tau_Q$ in the overdamped regime (Fig. \ref{fig5}(a)), along with only a slow (logarithmic) dependence on the initial value of the order parameter $\varphi(0)$ (hence, only logarithmic dependence on he temperature of the noise).

  It follows that, in accord with Eq. (\ref{solu}), the freeze-out time $\hat{t}$ obeys the relationship \cite{whz, whz2, whz3, lg2}
\begin{eqnarray}\label{that1}
\hat{t}\propto \sqrt{\eta\tau_Q}
\end{eqnarray}
in the overdamped case.

\section*{Underdamped and overdamped evolution}

For the underdamped case, Eq. (\ref{ode1}) can be  solved numerically. Fig. \ref{fig5} (b) depicts the rescaled solution $\tilde{\varphi}(\tilde{t})$ of Eq. (\ref{ode1}) where the second term $\eta\dot{\varphi}$ is neglected and $\tilde{\varphi}=\tau_Q^{1/3}\varphi$, $\tilde{t}= t/\tau_Q^{1/3}$. After this rescaling, the solutions corresponding to different quench timescales $\tau_Q$ exhibit a clear similarity.

Therefore, the freezeout time $\hat{t}$ obeys the relationship \cite{whz,lg2}:
\begin{eqnarray}\label{that2}
\hat{t} \propto \tau_Q^{1/3}
\end{eqnarray}
in the underdamped case.

\begin{figure*}[t]
{%
\includegraphics[clip,width=2\columnwidth]{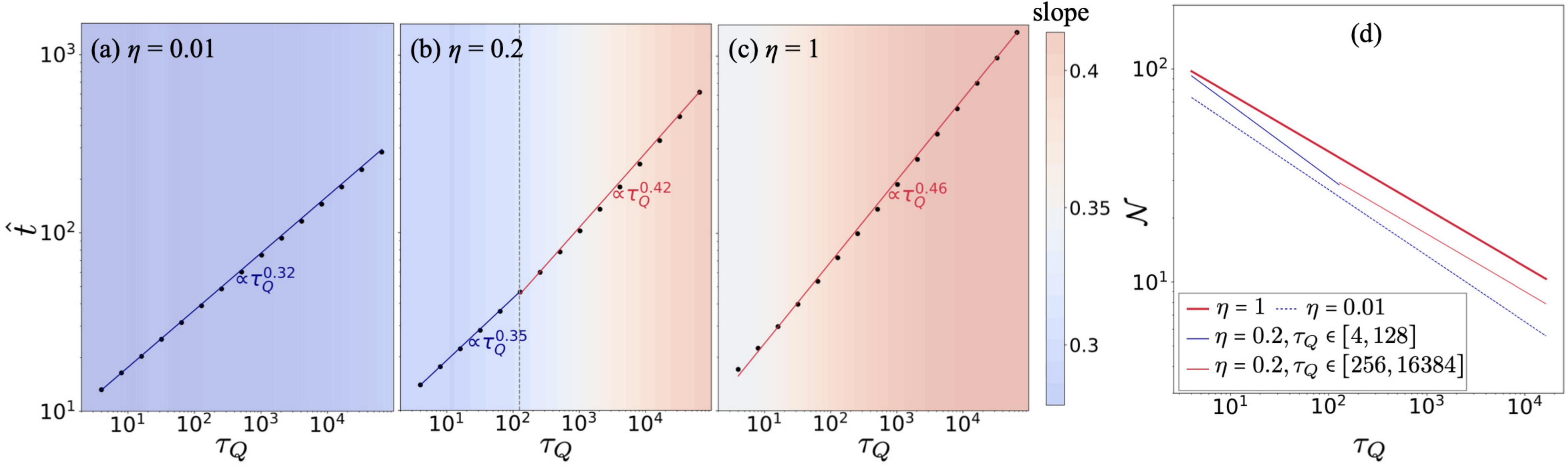} 
}
\caption{Transition of the freeze-out time scaling between the underdamped and overdamped regimes, obtained from the location of the inflection point of the solution to Eq. (\ref{ode1}). The freeze-out time $\hat{t}$ inferred from the first inflection point where $\ddot{\varphi}=0$ as a function of the quench timescale $\tau_Q$ for damping constants (a) $\eta = 0.01$, (b) $\eta = 0.2$, and (c) $\eta=1$, with $\varphi (0)=10^{-4}$. The color plot represents the slope of the log plot based on nearest neighbor points. For $\eta = 0.2$, a transition from the underdamped regime to the overdamped regime is observed as $\tau_Q $ increases. The dashed gray line represents the theoretical prediction  $\tau_Q = 1/\eta^3$ where the transition from underdamped to overdamped scaling occurs \cite{lg2}. (d) The number of defects $\mathcal{N}$ as the function of $\tau_Q$ for $\eta=0.01$ (dashed blue line), $\eta=0.2$ (solid blue line for $\tau_Q\in[4,128]$ and solid red line for $\tau_Q\in[128,16384]$), and $\eta=1$ (thick red line) with $\theta=10^{-8}$. $\mathcal{N}$ is obtained by numerically solving the full Langevin PDE (Eq. (\ref{langevin})) in (1+1) dimensions for a real scalar field.} 
\label{fig6}
\end{figure*}

Equation (\ref{langevin}) that dictates the evolution of the order parameter and is used to study formation of topological defects can have, in general, both first- and second-order time-derivative terms. As we have seen, purely overdamped and purely underdamped temporal evolution leads to different scaling of $\hat t$ and of the defect density. The question then arises: What is the scaling when both $\dot \varphi$ and $\ddot \varphi$ are present? 


Evolution generated by Eq. (\ref{langevin}) is---in the spirit of the KZM---expected to be overdamped when $\eta\,\dot \varphi > \ddot \varphi$ at the freeze-out time $\hat{t}$. In this regime, the relaxation time $\tau_{\dot \varphi} \simeq |\varphi/\dot\varphi|$ scales with $\epsilon$ as:
$
\tau_{\dot \varphi} \simeq \eta |\epsilon|^{-1} 
\simeq \eta\,\tau_Q \,|t|^{-1}\, 
$. In accord with \cite{whz}, one expects the size $\hat\xi$ of the domains of the new broken symmetry phase to be set at the time $\hat t$, when the time to (from) the phase transition is comparable to the relaxation timescale,
and the freeze-out of the field configuration occurs as a result of critical slowing down. This freeze-out condition,
$\tau_{\dot \varphi}(\hat t_{\dot \varphi}) = \hat t_{\dot \varphi}$
yields in this case;
\begin{eqnarray}
\label{eq:5}
\hat t_{\dot \varphi}  \simeq  (\eta\,\tau_Q)^{1/2}, \quad \hat\epsilon_{\dot \varphi}  \simeq  
\left(\frac{\eta}{\tau_Q}\right)^{1/2} \, .
\end{eqnarray}
The correlation length $\hat \xi$ which sets the stage for the defect formation is then:
\begin{equation}
{\hat\xi}_{\dot \varphi} \simeq 
\frac{\xi_o}{|\hat\epsilon_{\dot \varphi}|^{1/2}}
\simeq \xi_o \left(\frac{\tau_Q}{\eta}\right)^{1/4} \, .
\label{eq:7}
\end{equation}
 The density of the number of kinks is given by
\begin{equation}
{\cal N}_{\dot \varphi} \simeq \frac{1}{\hat\xi_{\dot \varphi}}
\propto \left(\frac{\eta}{\tau_Q}\right)^{1/4} .
\label{eq:7b}
\end{equation}

In the underdamped case,  $\ddot \varphi$ will dominate, and the order parameter reacts to the quench-induced changes in the effective potential on the timescale
$\tau_{\ddot \varphi} \simeq |\varphi/\ddot\varphi|^{1/2}$. Thus,
$
\tau_{\ddot \varphi} \simeq
\tau_o\,|\epsilon|^{-1/2} \, .
$
The freeze-out condition, 
$\tau_{\ddot \varphi}(\hat t_{\ddot \varphi}) = \hat t_{\ddot \varphi}$ 
yields in this underdamped regime:
\begin{eqnarray}
\hat t_{\ddot \varphi}  \simeq  (\tau_Q/\tau_o)^{1/3} , \quad \hat \epsilon_{\ddot \varphi}  \simeq  (\tau_o/\tau_Q)^{2/3} \, .
\label{eq:5b}
\end{eqnarray}
Consequently, the scaling of the characteristic correlation length with the quench time $\tau_Q$ is expected to change to
\begin{equation}
\hat\xi_{\ddot \varphi} \simeq 
\frac{\xi_o}{|\hat\epsilon_{\ddot \varphi}|^{1/2}}
\simeq \xi_o {\tau_Q}^{1/3} \, .
\label{eq:7b}
\end{equation}
Furthermore, the density of the number of kinks is given in this case by:
\begin{equation}
{\cal N}_{\ddot \varphi} \simeq \frac{1}{\hat\xi_{\ddot \varphi}}
\propto \left(\frac{1}{\tau_Q}\right)^{1/3} .
\label{eq:8b}
\end{equation}

We can therefore draw two related conclusions: (i) In the overdamped regime, the density of kinks should scale with
$\eta^{1/4}$, and should become viscosity independent in the underdamped case. (ii) Power-law dependence of the density of kinks
with the quench timescale should change from $\propto \tau_Q^{-1/4}$ in the overdamped case to $\propto \tau_Q^{-1/3}$ in the
underdamped case. The overdamped scalings should apply when the evolution is dominated by the first derivative 
($\eta\,\dot \varphi > \ddot \varphi$, i.e.,  $\eta/\tau_{\dot\varphi} > 1/\tau_{\ddot\varphi}^2$) at the instant when topological defects freeze-out.
This will happen for:
$
|\hat\epsilon_{\dot\varphi}| > |\hat\epsilon_{\ddot\varphi}| \, ,
$
or---using Eq. (\ref{eq:5}) and (\ref{eq:5b})---when:
$$
\eta^{3} > 1/\tau_Q \,,
\label{eq:10}
$$
as discussed in \cite{lg2}.

We have seen that  the freeze-out time $\hat{t}$ can be determined directly from the location of the inflection point of the solution of Eq. (\ref{ode1}). Analyzing the solution of the ODE without assuming either the overdamped or underdamped limits reveals the transition from the scaling behavior of the overdamped regime to that of the underdamped regime. We rely on the observation that the freeze-out time $+\hat{t}$ is the moment when the solution $\varphi(t)$ begins to move rapidly toward the broken symmetry potential minima after $t=0$. This happens at the inflection point when the second time-derivative $\ddot{\varphi}=0$ for the first time after $t=0$. In Fig. \ref{fig6}, $\hat{t}$ identified as such inflection point is plotted as a function of the quench timescale $\tau_Q$ for (a) $\eta=0.01$, (b) $\eta=0.2$, and (c) $\eta=1$, with $\varphi (0)=10^{-4}$.

In Fig. \ref{fig6} (a-c), the color plots reflect the slope of the log plot based on nearest neighbor points. It is found that $\hat{t}\propto \tau_Q^{0.46}$ for $\eta=1$ and 
$\hat{t}\propto \tau_Q^{0.32}$ for $\eta=0.01$; Both closely align with the relations given by Eqs. (\ref{that1},\ref{that2}) respectively, as predicted in \cite{lg2}. The color plot reveals a transition from the underdamped regime to the overdamped regime as $\tau_Q$ increases for $\eta=0.2$. We have that $\hat{t}\propto \tau_Q^{0.35}$ for  $\tau_Q\in [4,128]$ and $\hat{t}\propto \tau_Q^{0.42}$ for $\tau_Q\in [128, 16384]$. The gray dashed line represents the theoretical prediction, $ \tau_Q=1/\eta^3$, where the gradual transition between the overdamped ($\hat t \sim \tau_Q^{\frac 1 2}$) and underdamped ($\hat t \sim \tau_Q^{\frac 1 3}$) regime is expected \cite{lg2}. 

Fig. \ref{fig6} (d) shows the number of defects $\mathcal{N}$ created by phase transitions as a function of $\tau_Q$. This result is obtained by numerically solving the full PDE (Eq. (\ref{langevin})) in (1+1) dimensions with $\theta=10^{-8}$ (Appendix). In the underdamped case with $\eta=0.01$, the number of defects $\mathcal{N}\propto \tau_Q^{-0.31}$ (dashed blue line), while $\mathcal{N}\propto \tau_Q^{-0.27}$ in the overdamped case with $\eta=1$ (thick red line). For $\eta=0.2$, we observe a change in the scaling  of the number of defects. We have $\mathcal{N}\propto \tau_Q^{-0.34}$ for small quench timescales $\tau_Q \in [4,128]$ and $\mathcal{N} \propto \tau_Q^{-0.27}$ for large quench timescales $\tau_Q\in [128,16384]$.
 The scaling of the number of defects (${\cal N} \sim {\tau_Q}^{- \frac 1 3}$ to ${\cal N} \sim {\tau_Q}^{-\frac 1 4}$) corresponds to the change in the scaling of the freeze-out time $\hat{t}$ from the underdamped to the overdamped regime as $\tau_Q$ increases. This is anticipated by the KZM based on the analysis of the consequences of critical slowing down \cite{whz, lg2}, but it can be predicted solely by solving simple ODE (Eq. (\ref{ode1})).\\

 \begin{figure}[t]
 \centering
{%
\includegraphics[clip,width=0.9\columnwidth]{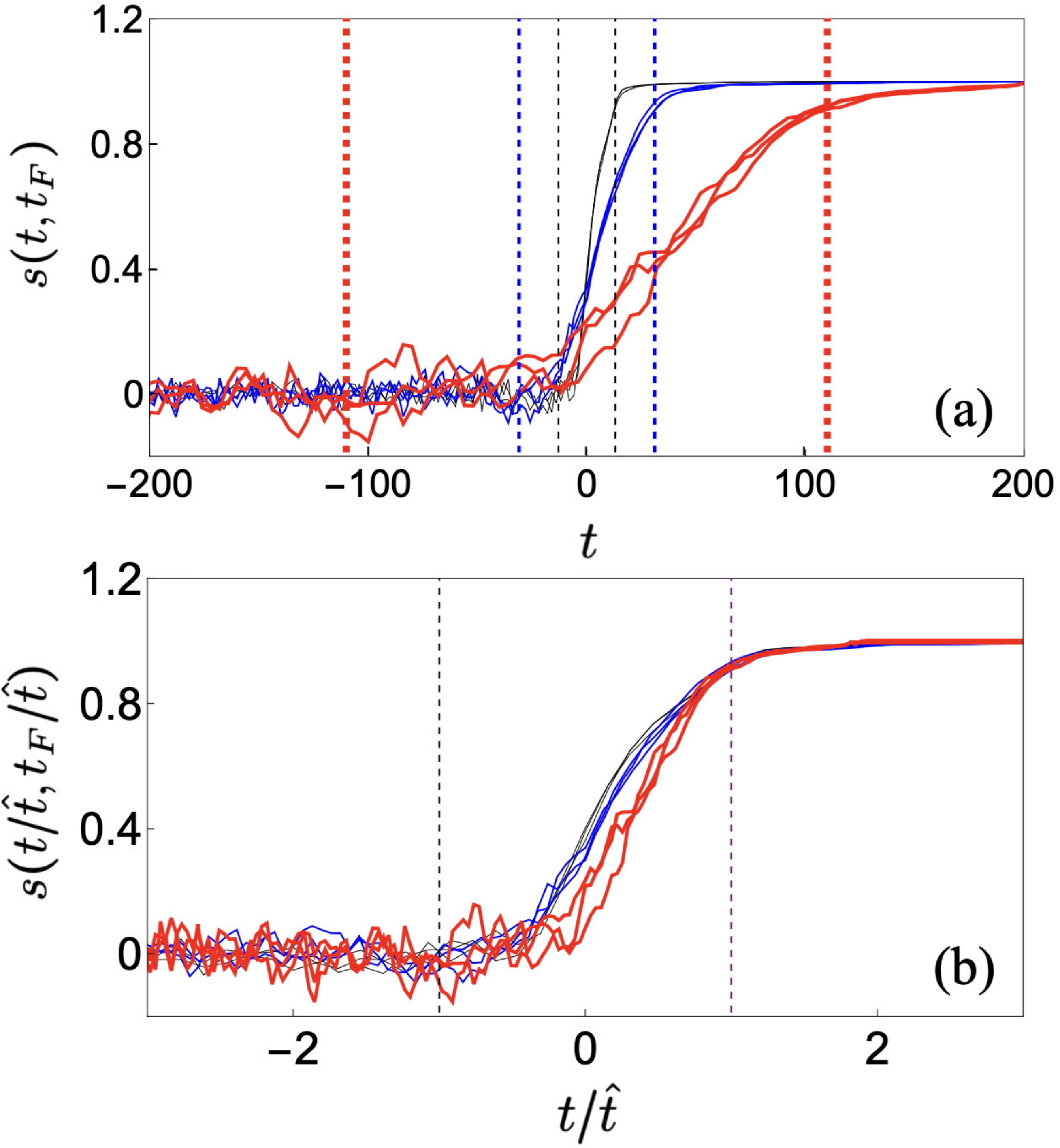} 
}
\caption{ Gradual evolution of $\Phi (x,t)$ toward its final configuration, obtained by numerically solving the full Langevin PDE (Eq. (\ref{langevin})) in (1+1) dimensions for a real scalar field. The time-correlation function $s(t,t_F)$ (Eq. (\ref{scalar})) for $\theta=10^{-2}$, and three symmetry-breaking evolutions for each of three quench times: $\tau_Q=8$ (thin, black), $\tau_Q=64$ (blue), and $\tau_Q=512$ (thick, red). The locations of topological defects settle by $t_F=200$, and the correlation function approaches the asymptotic value of 1 when $\Phi (x,t)$ and $\Phi (x,t_F)$ are obtained from the same run---otherwise $s(t,t_F) \approx 0$. The correlation function $s(t,t_F)$ in (a) is plotted as the function of $t$. The vertical dashed lines indicate $\pm \hat{t}$ for each $\tau_Q$, while (b) $s(t,t_F)$ is plotted as a function of rescaled time $t/\hat{t}$, where the purple line indicates $\pm \hat{t}$ for this case. Now the plots for different $\tau_Q$ nearly overlap, confirming the universality predicted by the KZM.}
\label{fig7}
\end{figure}

\section*{Freeze-out of the spatial structure at $\hat t$}

We now turn to the question: How is the timescale $\hat{t}$ imprinted on the spatial structure of $\Phi(x,t)$? To address this, we consider the equation obtained by removing the time dependence and the noise term from Eq.~(\ref{langevin}). The resulting equation for the spatial dependence of $\Phi(x,t)$ is:
\begin{eqnarray}\label{spatial}
\nabla^2 \phi (\mathbf{x}) -\partial_{\phi} V(\phi) =0 .
\end{eqnarray}
It is solved by
\begin{eqnarray}\label{jacobi}
\phi (x)=\phi(0)\mbox{cd}\left(\frac{i\sqrt{\phi(0)^2-2\epsilon}x}{2},-\frac{\phi(0)^2}{\phi(0)^2-2\epsilon}\right)
\end{eqnarray}
in 1-dimensional case where $\phi(0)=\phi(x=0)$, $\phi'(0)=0$ and $\mbox{cd}$ represents Jacobi elliptic function. Eq. (\ref{jacobi}) can be approximated by $\phi (x)=\phi (0) \cos \sqrt{\epsilon/2}x$ for small $|\phi (0)| \ll \sqrt{\epsilon}$ when $\epsilon >0$. It exhibits spatial periodicities related to $1/ \sqrt \epsilon $.  Of particular interest is the periodicity $1/\sqrt{\hat{\epsilon}}$ that emerges for small $\phi$ near $+\hat{t}$ when the order parameter $\Phi(x,t)$ begins to grow from the small fluctuations bestowed during the $[- \hat t, +\hat t]$ interval by the random walk $\vartheta(x,t)$ in the Langevin equation Eq. (\ref{langevin}), and rises to the broken symmetry equilibrium $\sqrt{\epsilon}$. Focusing on its spatial part, we can think of the local $\phi(x)$ as a consequence of the random walk filtered by the spatial component of the Langevin equation, Eq. (\ref{spatial}). That filtering attempts to impose periodicities (set by the ``spring constant’’ $\sim  \hat \epsilon $) of the Jacobi function (which represents the solution of the ``physical pendulum’’). That spatial period is exactly $ \sim 1/\sqrt {\hat \epsilon} $, as for small values of $\phi$ the effect of the nonlinearity in Eq. (\ref{spatial}) can be neglected. This yields
\begin{eqnarray}
\hat{\xi} \sim   \begin{cases}
                        (\tau_Q/\eta)^{1/4}, \qquad \text{  (overdamped case)} \\
                        \tau_Q^{1/3}, \qquad \qquad \quad \text{(underdamped case)}
                    \end{cases}.
\end{eqnarray}
This structure is both imprinted and erased on $\Phi (x,t)$ by the combination of the filtering and random walk due to the noise term in the interval $[- \hat t, +\hat t]$. It firms up only after $+\hat t$, as the order parameter settles into the potential minima associated with broken symmetry.

The evidence of its gradual accumulation, which for \( t > +\hat{t} \) leads to the formation of well-defined topological defects, can be seen in the time-correlation function (Eq. (\ref{scalar})) with the final configuration \( \Phi(x, t_F) \), evaluated at \( t_F \), the time after the system has settled to its post-transition state following \( +\hat{t} \). It is a measure of the proximity of evolving $\Phi (x,t)$ with the final configuration:
\begin{eqnarray}\label{scalar}
 s(t,t_F) = \frac{\sum_{x}\Phi (x,t)\Phi (x,t_F)}{\sqrt{\sum_{x}\Phi (x,t)^2}\sqrt{\sum_x \Phi (x,t_F)^2}}.
 \label{scalar}
\end{eqnarray}
In Fig. \ref{fig7} (a), we see that $s(t,t_F)$ begins to rise already somewhat before the critical point is traversed, but after $-\hat t$, and reaches its equilibrium value shortly after $+ \hat t$ when, according to \cite{whz}, the basic structure of the broken symmetry state (including the location of the defects) is determined. This confirms the adiabatic-impulse-adiabatic parsing of the dynamics of symmetry breaking quenches. Furthermore, panel~(b) presents $s (t, t_F)$ as a function of the rescaled time \( t/\hat{t} \), demonstrating that the curves for different \( \tau_Q \) closely overlap, in agreement with the universality  predicted by the KZM.

We note that, while our discussion was focused on the Langevin equation for Landau-Ginzburg and Gross-Pitaevskii models of the phase transition in 1+1 dimension, periodicities related to the solutions of Eq. (\ref{spatial}) will arise as long as the Laplacian operator $\nabla^2$ is involved, regardless of the number of spatial dimensions. This appears to be the case in systems where the sonic horizon motivation for the KZM is difficult to invoke, as the propagation of the front of the broken symmetry involves diffusion rather than sound, or is a subject to other constraints (e.g., involves a conserved order parameter).

\section*{Discussion}
The KZM predicts consequences of nonequilibrium phase transitions in diverse settings including condensed matter \cite{exp, dorner,  bec,bec2,reichhardt, reichhardt2,ralf,adolfo,anderson, jacek, mithun, bayocboc,wit,carmi,monaco,sadler,weiler,chiara,benni,ulm,pyka,chae,lin,griffin,dona,mono,choma,navon,rysti,saito,cyn, vortex3, moss},  chemistry \cite{nik,nik2}, quantum computing \cite{damski,zoller,dziarmaga,dziarmaga2, polkovnikov,polkovnikov2,lukasz, qc,qc2,qc3,qc4, qc5,qc6, bando,gardas,universal,anders,ralfqc},  cosmology \cite{kibble,kibble2,whz,whz2,whz3, hindmarsh, cosmology,cosmology2,cosmology4, beilok}, and in other systems where a spatially uniform pretransition state is forced to undergo symmetry breaking.  While creation of topological defects and other excitations in nonequilibrium phase transitions is the best known application, the KZM has been successfully used to study structure formation in transitions between steady states in systems such as Rayleigh-Benard instability \cite{casado2006testing}. 

The cornerstone of the KZM is the conjecture that the size of the domains that choose the same broken symmetry corresponds to the equilibrium correlation (or healing) length at the instant when the relaxation rate of the system equals to the speed with which the relevant (e.g., thermodynamic) potential is changing \cite{whz}. This leads to the KZM scaling that predicts the dependence of the size of domains in the post-transition state on the quench rate. It is by now widely tested in both laboratory and numerical experiments and verified whenever the assumptions that lead to the KZM hold.

The question we considered is why the KZM works so well. To address it we examined a paradigmatic model---a field theory with the time-dependent Landau-Ginzburg potential---where the KZM is known to hold, as the density of topological defects (e.g., ``kinks’’) generated by quench  follows the KZM scaling \cite{lg,lg2}. 
To understand why the KZM works so well we deconstructed the PDE used to study the KZM and examined the role of its components, starting with the ODE responsible for temporal evolution. 

The advantage of this ``piecewise’’ approach is evident: Temporal ODE is easy to integrate. Moreover, in the overdamped case it reduces to the analytically solvable Bernoulli equation. The time-dependent evolution of the order parameter exhibits features---e.g., marks the instants $\pm \hat t$---that have been identified (on the basis of heuristic arguments) as the cornerstones of the scaling predictions of the KZM. These two instants mark the end and the beginning of the time intervals where the order parameter can adiabatically follow the potential imposed externally. Between these two instants---in the time interval delineated by $-\hat t$ and $+\hat t$---the reaction time of order parameter of the system is too slow to allow for adiabatic evolution.
In the pre-transition period that ends at the critical point at $t=0$ the order parameter is buffeted by noise. It is especially vulnerable to the noise after $t=-\hat t$ is traversed, as the restoring force of the quadratic part of the potential is too weak to reimpose the pre-transition symmetric equilibrium state. Thus, after the critical point is traversed,  the instability at $t>0$ begins to be dominate the evolution as the order parameter starts with a small but non-negligible initial (i.e., at $t \simeq 0$) value imposed by the noise-driven random walk. 

The subsequent evolution in the ``upside down’’ harmonic oscillator potential amplifies the $\Phi(x,t=0)$ configuration. The order parameter that was ``primed’’ by fluctuations will start to evolve towards one of the broken symmetry minima. 
The expected values of $\Phi(x,t=0)$ are easy to estimate but are difficult to calculate precisely, as the spatial part plays the role. Fortunately, their effect on $\hat t$---the crucial element of the KZM---is only logarithmic (hence, negligible).  

This rapid (superexponential) evolution for $t>0$ slows down only at $t \gtrsim +\hat t$, as the local broken symmetry minimum is reached. The structure of the post-transition configuration of the order parameter---including, in particular, topological defects---is then seeded by the fluctuations. Topology stabilizes defects after $+\hat t$, although annihilation is still possible. Their density is largely determined by the size of the broken-symmetry domains at $+\hat t$. 

Spatial part of Eq. (\ref{langevin}) plays a key supporting role, as it exhibits spatial periodicity related to $\hat t$ which, at freeze-out time, corresponds to $\hat \xi$. Thus, fluctuations pass through the filter that favors structures with the approximate spatial periodicity of $\hat \xi$. This reasoning applies whenever the homotopy group allows for existence of topological defects \cite{kibble}, and whenever the spatial part of the evolution includes the Laplacian operator.
When the order parameter is complex, phase ordering can be rapid, but in a closed loop---at least in the 1D case we investigated---it will leave winding numbers with the expectation values that can be estimated from the quench rate \cite{whz}. The scaling of the expected winding numbers depends on the rate of the transition and the size of the loop: When the diameter of the loop is large compared to $\hat \xi$, the winding number will be given by the square root of the number of defects that are expected to form inside that loop. When the loop has diameter comparable or smaller than $\hat \xi$, the probability of trapping a winding number is given by the probability of finding a single defect within the loop of that size \cite{zurek2013topological}.

Previous studies of the KZM have primarily focused on testing the power-law scaling it predicts. This work highlights the role of the dynamics of the order parameter, which should be directly accessible in experiments, and could provide important insight into the foundational conjecture underlying the KZM. The time evolution of the order parameter, such as magnetization, polarization, or BEC density, is often an experimentally observable quantity. Remarkably, as demonstrated in this paper, these dynamics of the order parameters can be described by simple, often exactly solvable ODEs. Symmetry breaking dynamics in second order phase transitions has been so far studied indirectly, primarily through the quench rate dependence of the density of topological defects---relics of such phase transitions. Our results suggest a possibility of a more direct approach---following the evolution of the order parameter. This is in a sense an approach complementary, yet (as we have seen) closely tied to the emergence of defects. It is certainly possible to monitor growth of the order parameter (see, for example, analysis of the time-dependent growth of the condensate density in \cite{liu}, and experimental results such as \cite{miesner, hugbart, ritter}). It remains to be seen whether dedicated experiments can be done with sufficient temporal resolution and in systems that are uniform enough so that the spurious spatial gradients do not obscure the growth of the order parameter predicted by the relevant ODE. In addition to these implications for  potential future tests, the study of order-parameter dynamics presented in this paper has recently inspired the use of machine learning to predict the locations of topological defects \cite{ML}.

\begin{acknowledgements}
We thank John Bowlan for helpful discussions. F.S. acknowledges support from the Los Alamos National Laboratory LDRD program under project number 20230049DR and the Center for Nonlinear Studies under project number 20250614CR-NLS.
\end{acknowledgements}

\bibliographystyle{unsrt}
\bibliography{pnas-sample}

\appendix

\renewcommand{\theequation}{S.\arabic{equation}}
\setcounter{equation}{0}

\section*{Numerical methods}

We employed the 4th-order Runge-Kutta method to  solve the full Langevin PDE (Eq. (1)) and the Gross-Pitaevskii equation (Eq. (5)) numerically. For the Langevin equation, we chose the system size  $L=2048$ with 4096 grid points. The number of defects $\mathcal{N}$ shown in Fig. 6  (d) is obtained by numerically solving the full PDE (Eq. (1)) in (1+1) dimensions 15 times (i.e., 15 realizations of random noise), starting the time evolution at  $\epsilon=-1$ and concluding it at 
$t=32768$ (i.e., $\epsilon=1$ for $\tau_Q=16384$). For the stochastic Gross-Pitaevskii equation, we chose the system size  $L = 30$ and 300 grid points to generate the plot in the Fig. 4.

\section*{Correlation length}

In addition to the method used in the main text, we can also estimate the correlation length as follows.   We consider  the following equation that excludes the time dependence and the noise term from the Langevin equation Eq. (1) in the main text, and by introducing a delta-function source $-\delta (x)$ at the origin:
\begin{eqnarray}
\nabla^2 \phi (\mathbf{x}) -\partial_{\phi} V(\phi) =-\delta (\mathbf{x}).
\end{eqnarray}
We assume $\phi (\mathbf{x})=\phi_0 (\mathbf{x}) +\delta \phi (\mathbf{x})$ where $\nabla^2 \phi_0 (\mathbf{x}) -\partial_{\phi_0} V(\phi_0)=0$ while the perturbation $\delta \phi$ obeys
\begin{eqnarray}
\nabla^2 \delta \phi (\mathbf{x})+\frac12  \epsilon (t)\delta \phi (\mathbf{x})=-\delta (\mathbf{x}).
\end{eqnarray}
Here the higher-order terms of $\delta \phi (\mathbf{x})$ are discarded. The solution of this equation takes the Ornstein-Zernike form:
\begin{eqnarray}
\delta \phi (\mathbf{x}) \sim |\mathbf{x}|^{-(d-1)/2}\exp (-|\mathbf{x}|/\xi)
\end{eqnarray}
in $d$-dimensional space and $\xi \sim 1/\sqrt{\epsilon}$. This yields
\begin{eqnarray}
\hat{\xi} \sim   \begin{cases}
                        (\tau_Q/\eta)^{1/4}, \qquad \text{  (overdamped case)} \\
                        \tau_Q^{1/3}, \qquad \qquad \quad \text{(underdamped case)}
                    \end{cases}
\end{eqnarray}
where
\begin{eqnarray}
\hat t_{\dot \varphi}  \simeq  (\eta\,\tau_Q)^{1/2}, \quad \hat\epsilon_{\dot \varphi}  \simeq  
\left(\frac{\eta}{\tau_Q}\right)^{1/2} \, 
\end{eqnarray}
in the overdamped regime, and
\begin{eqnarray}
\hat t_{\ddot \varphi}  \simeq  (\tau_Q/\tau_o)^{1/3} , \quad \hat \epsilon_{\ddot \varphi}  \simeq  (\tau_o/\tau_Q)^{2/3} \, 
\end{eqnarray}
in the underdamped regime from Eqs. (11, 14) in the main text.

\section*{Gradual convergence of domain sizes towards $\hat{\xi}$}

\begin{figure}[t]
 \centering
{%
\includegraphics[clip,width=1\columnwidth]{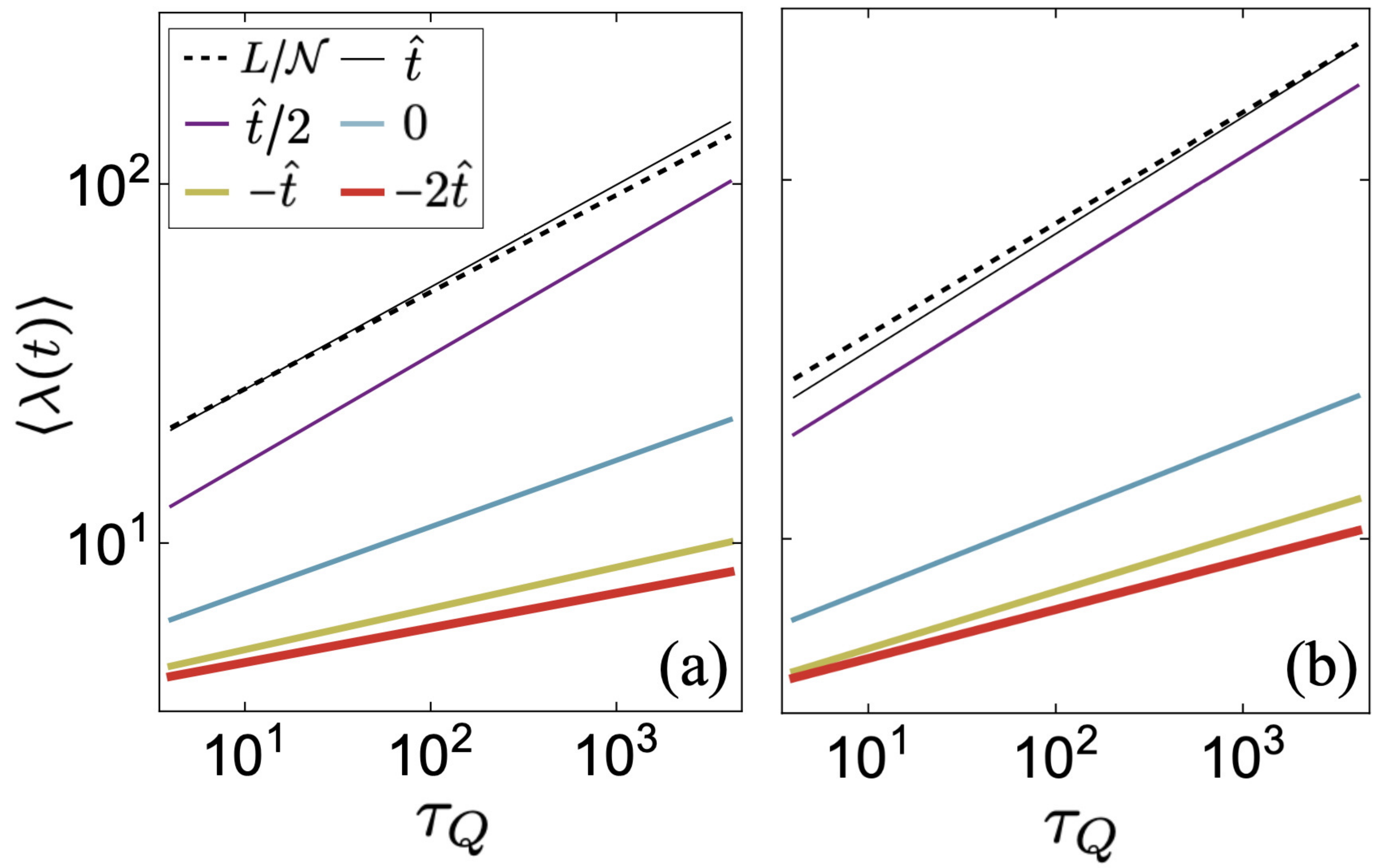} 
}
\caption{ \textbf{Gradual accumulation toward the final configuration of $\Phi$.} $\langle \lambda (t) \rangle $ as a function of $\tau_Q$ for   $\eta=1$ (a) and for  $\eta=0.01$ (b). From thick red to thin black line, $t=-2\hat{t}, -\hat{t}, 0, \hat{t}/2, \hat{t}$ respectively. The dashed black lines represent $L/\mathcal{N}$ (i.e., the average domain size) from Fig. 6 (d). $\theta=10^{-8}$.}
\label{fig8}
\end{figure}

In addition to the correlation $s(t,t_F)$ with the final configuration shown in Fig.~7, we can also examine the evolution of the expected domain size to reveal the gradual accumulation of the order parameter. We identify the expected domain size with half of the spatial period, $\langle \lambda (t)\rangle$, obtained from the Fourier transform:
\begin{eqnarray}
\frac{1}{\langle \lambda (t)\rangle}=\frac{1}{\mathcal{C}}\sum_k \frac{2k}{L} \tilde{\Phi} (k,t) \tilde{\Phi}^{*} (k,t)
\end{eqnarray}
where $ \tilde{\Phi} (k,t)=\sum_{x}\Phi (x,t) e^{-i 2\pi  k x/L}$, the normalization $\mathcal{C}= \sum_k \tilde{\Phi} (k,t) \tilde{\Phi}^{*} (k,t)$, and $L=2048$ is the system size. Here, we consider  half the spatial period $L/2k$ since the full period of the oscillation would produce a pair of kinks. Fig. \ref{fig8}  shows $\langle \lambda (t) \rangle $ as a function of $\tau_Q$ from $t=-2\hat{t}$ to $\hat{t}$ for the overdamped case with $\eta=1$ (a) and for the underdamped case with $\eta=0.01$ (b). The figure is obtained by averaging the results of 15 numerical simulations of Eq. (1). As time approaches $\hat{t}$, the behavior governed by the periodicity from the spatial component of the Langevin equation begins to dominate over that driven by the noise. The scaling of $\langle \lambda (t)\rangle$ eventually aligns closely with $L/\mathcal{N}$ (i.e., the average domain size) at $t=\hat{t}$ where $\mathcal{N}$ is the number of defects from Fig. 6 (d). 

\end{document}